\documentstyle[12pt]{article}
\date{}
\title{On propagation of photon in a medium} 
\author{V.A.Maisheev \thanks{E-mail maisheev@mx.ihep.su} \\
{\it Institute for High Energy Physics, 142284, Protvino, Russia }}

\begin{document}
\maketitle
\def\arcctg{\mathop{\rm arccot}\nolimits} 
\def\ch{\mathop{\rm ch}\nolimits}
\def\sh{\mathop{\rm sh}\nolimits} 
\def\Im{\mathop{\rm Im}\nolimits}
\def\Re{\mathop{\rm Re}\nolimits}
\def\sign{\mathop{\rm sign}\nolimits}
\begin{abstract}

The equations for variation of the Stokes parameters and intensity
of photons propagating in a medium, whose optical properties may be
described by the permittivity tensor, are derived. 
Classification of  different cases of photon propagation
is suggested on a basis of these equations.
\end{abstract}

\section{ Introduction}
It is well known that polarized photons propagating in  an optically
active medium change initial polarization state \cite{LL,AG}. It has 
 important meaning in a lot of fields of physics.
We have pointed to some  existing problems:   

a) propagation effects in a magnetosphere of pulsars \cite{Puls1, Puls2};

b) results of the circular polarization measurements of extragalactic
radio sources \cite{Cir1} and \cite{Cir2} (in our Galaxy);

c) cosmic birefringence (see \cite{CB} and literature therein).

d) problems of propagation x-rays \cite{XX} and high energy $\gamma$-quanta
\cite{C,MV} in single crystals and creation of the polarimeters
for hard photon beams;    

e)experimental observation of the birefringence in 
an electromagnetic vacuum \cite{Bak,A};

f)investigations of propagation of $\gamma$-quanta 
in laser waves \cite{KS,MV2}  with
the aim of polarization control on future $\gamma \gamma$-colliders.

Besides, the possibility of propagation  polarization
phenomena  of the radiation emitted from quark-gluon plasma 
must not be ruled out.
 
Various phenomena (birefringence, Faraday rotation and etc.) of 
the visible light are well known and described in literature \cite{LL,AG}.
In \cite{AG} the detailed consideration of these processes was
carry out in anisotropic and gyrotropic media. However,
the description of propagation was realized on a basis  of the
polarization states, whereas representation in terms of density matrices   
is more adequate in some cases. A parallel with well known ones,
the new cases of propagation of photons were investigated in recent
papers \cite{KS,MV2,MV3}. In this connection the problem of
classification of these cases  and its  completeness arises.

 On the other hand, the description of the optical properties of a medium 
with the help of the permittivity tensor is the universal and 
traditional method.
This method may be used in  wide energy ranges of propagating  photons.
The proof of validity of the method for $\gamma$-quanta and samples of
its using one can find in \cite{LL1,TM,BT}. Using of permittivity tensor   
presumes the linear response of media on perturbations. It is a very good  
assumption for many cases (however, see \cite{ND}).

 The aims of this paper are: firstly, the derivation of  equations,
describing  all the cases of propagation of photons in media,
whose optical properties may be described with the help of the
permittivity tensor and, secondly, creation on a basis of these equations
classification of the various cases of photon propagation.  
Our consideration is based on some suggestions such as smallness
of the permittivity tensor components in compare to unit and straight line
motion of the photons. It is not the most general case, of course, but
the most interesting one for practical applications.

It should be noted, that the problems of propagation of the photon beam
were considered in some papers \cite{LS,MCB,BF}. For example, in \cite{BF} 
the equations of the general type for propagation process were
obtained.
However, it seems that their using in specific applications is difficulty,
because the physical nature of the parameters in the equations is unknown.

\section{Equations}
 We write the equations of the electromagnetic field in a medium
in the following form \cite{LL,AG}:
\begin{eqnarray}
rot {\bf{B}} = {{1}\over {c}} {{\partial {\bf{D}}}\over {\partial t}}
 \, ,\; \; \; div{\bf{D}} =0 \, , 
  \nonumber \\
 rot {\bf{E}} = - {{1}\over {c}} {{\partial {\bf{B}} } \over {\partial t}} 
\, , \; \; \; div {\bf{B}} =0 \, ,
\label{1}
\end{eqnarray} 
where
$ {\bf{E}}$  is the intensity of electric field and ${\bf{D}}$ and ${\bf{B}}$ 
are the electric and magnetic induction vectors,
t is the time, c is the speed of light. All the properties of the medium 
 are reflected
in the relation between ${\bf{B}}, {\bf{E}}$ and ${\bf{D}}$.
Eqs.(\ref{1}) would  suffice  to  describe  the   
propagation of  photons
in a medium and such a property as the intensity of magnetic field is not  
needed \cite{LL, AG}.  
We represent the
relation between ${\bf{D}}$ and ${\bf{E}}$ in the form
\begin{equation}
 D_i(\omega)=\varepsilon_{ij}E_j(\omega)\, , \; \; \; (i,j = 1,2,3 )\, ,
\label{2}
\end{equation} 
where $ \varepsilon_{ij}= \varepsilon'_{ij} + i\varepsilon''_{ij}$
is the complex permittivity tensor and 
$\omega$  is the frequency of the photon beam propagating in the medium.

We consider here only monochromatic photon beam, which propagates in 
defined direction. In this case conveniently to use in calculations
the inverse $\eta_{\alpha \beta}$ of the tensor $\varepsilon_{ij}$. 
In Cartesian coordinate system with one axis oriented along the direction
of photon motion this tensor can be represented as two-dimensial tensor
$(\alpha=1,2, \beta=1,2)$ because of perpendicularity of ${\bf{D}}$ to the 
photon wave vector. The obtained below equations are written  in this
coordinate system. 

In our consideration we assume that the photon beam move along straight line.
It means that in transverse directions (relative to motion of the beam)
a medium is uniform (\cite{LL}). In the most of practical cases it is 
true with very high accuracy.

Knowing the tensor $\eta_{\alpha \beta}$, one can find the refractive indices
of propagating photons \cite{AG,MV2,MV3}:
\begin{equation}
\tilde n^{-2} 
=(\eta_{11}+\eta_{22})/2 \pm 
\sqrt{(\eta_{11}-\eta_{22})^2/4 + \eta_{12}\eta_{21} }
\, ,
\label{3}
\end{equation}
Thus, in the general case the photon beam  
propagates through the
medium as the superposition of two electromagnetic waves with different
refractive indices. 
In the general case the refractive indices are  complex values.

 One can describe the polarization state either (of the two) wave  by 
the use of the Stokes parameters $X_i,\, Y_i \, (i =1-3)$ ($X_i$-values
correspond to one wave and $Y_i$ to another). These parameters 
are determined by the following relations \cite{MV2,MV3}
\begin{eqnarray}
X_1, Y_1 ={\kappa + \kappa^{*} \over 1+\kappa \kappa^{*}} \, , 
\label{4} \\
X_2, Y_2 = {i(\kappa - \kappa^{*}) \over 1+ \kappa \kappa^{*}} \, , 
\label{5} \\
X_3, Y_3 = {\kappa \kappa^{*} -1 \over 1 +\kappa \kappa^{*}} \, .
\label{6}
\end{eqnarray}
where $\kappa$ is the ratio of components of electric induction vector 
${\bf{D}}(D_1,\, D_2,\, 0)$ :

\begin{equation}
{D_1\over D_2} =\kappa = {{\tilde n^{-2} - \eta_{22}}\over {\eta_{21}}} =
{{|D_1|}\over {|D_2|}}e^{i\delta},
\label{7}
\end{equation} 
where $\kappa$ calculated with refractive index $\tilde{n_1},\,(\tilde{n_2})$ 
corresponds to $X_i\,(Y_i)$-parameters (denote the $\kappa$-values 
correspondingly as $\kappa_1$ and $\kappa_2$)  and 
$\delta$ is the phase shift between $D_1$ and $D_2$. 

This ratio
$\kappa$ can be reduced to zero or to the form $\kappa = i\rho$
(since $|D_1| |D_2| \sin \delta =b_1 b_2$, where $b_1$ and $b_2$ are the
semiaxes of the ellipse and $|\rho| = b_1 / b_2$ \cite{LL})  by the
rotation of the coordinate system around  the photon wave vector 
(the third  axis is constantly aligned with the wave vector). The first case 
corresponds to the propagation of a linearly polarized wave and the
second case corresponds to an elliptically polarized wave; in addition,
$\rho > 0 (\rho <0)$ corresponds to left (right) - hand polarization of
$\gamma$-quanta. 

From our consideration follows the important conclusion that these
two waves  (named as electromagnetic normal waves \cite{AG}) with  
determined refractive indices and polarization states are the
eigenfunctions of problem and any monochromatic electromagnetic 
wave propagating in the medium in defined direction is superposition 
of these two waves. 

 One can obtain the following relations:
\begin{equation}
\kappa_1 \kappa_2= -{\eta_{12} \over \eta_{21}},
\label{8}
\end{equation}
In the case of symmetric $\eta_{\alpha \beta}$-tensor, the product 
$\kappa_1 \kappa_2= -1$ and $X_1=-Y_1,\, X_2=Y_2, X_3=-Y_3$.
Another important case take a place when $\eta_{12}+\eta_{21}=0$ 
(but $\eta_{12} \ne 0$) and then 
$\kappa_1 \kappa_2 =1$, $X_1=-Y_1,\,X_2=Y_2,\, X_3=-Y_3$.    
However, in general the absolute values of circular and linear
polarizations for both the normal waves are not equal in between
(i.e. $|X_2| \ne |Y_2|,\, X_1^2+X_3^2 \ne Y_1^2+ Y_3^2 $).

Now we consider the case of photon propagation in the medium, whose
optical properties described by the tensor $\eta_{\alpha \beta}$ at condition
$\eta_{12}+\eta_{21}=0$, ($\eta_{12} \ne 0$). This case was investigated
in \cite{MV2}.
The variations of intensity and Stokes parameters
one can write in the following form:

\begin{eqnarray} 
J_{\gamma}(x)= J_1(x)+J_2(x)+2J_3(x) \, , 
\label{9} \\
\xi_1(x)= (X_1 J_1(x) + Y_1 J_2(x)+ p_1 J_3(x))/J_{\gamma}(x)\, ,
\label{10} \\
\xi_2(x) =(X_2 J_1(x) + Y_2 J_2(x) + p_2 J_4(x))/J_{\gamma}(x) \, , 
\label{11} \\
\xi_3(x)=(X_3 J_1(x) + Y_3 J_2(x) +p_3 J_4(x))/J_{\gamma}(x) 
\, ,\label{12}\\
p_1={2 \over X_1},\,p_2=-{2X_3\over X_1},\, p_3={2X_2\over X_1} \, ,
\label{13}
\end{eqnarray}
where  $J_{\gamma}, \xi_1, \xi_2, \xi_3$  are the intensity and 
Stokes parameters of the photon beam on
the distance equal to x.  The partial intensities $J_i(x),
(i=1-4) $ have the following form:
\begin{eqnarray}
J_1(x)=J_1(0)\exp(-2\Im(\tilde n_1)\omega x/c) \, ,
\qquad \qquad \qquad \label{14} \\
J_2(x)=J_2(0)\exp(-2\Im(\tilde n_2)\omega x/c) \, ,
\qquad \qquad \qquad \label{15} \\
J_3(x)=\exp(-\Im(\tilde n_1 +\tilde n_2 ) 
\omega x/c)\{J_3(0)\cos(\Re(\tilde n_1 -
\tilde n_2)\omega x/c) - \\ 
  J_4(0) \sin(\Re(\tilde n_1 - \tilde n_2) \omega x/c)\} 
 \,,\label{16} \nonumber \\
J_4(x)=\exp(-\Im(\tilde n_1+ \tilde n_2) \omega x/c)\{J_3(0) 
\sin(\Re(\tilde n_1 -
\tilde n_2) \omega x/c) + \\
J_4(0)\cos(\Re(\tilde n_1- \tilde n_2)\omega x/c)\} 
\,. \label{17} \nonumber
\end{eqnarray} 
The initial partial intensities are defined from the following relations:
\begin{eqnarray}
J_1(0)= {{1-X_1\xi_1(0) +X_2\xi_2(0)+ X_3\xi_3(0)}\over { 2(X_2^2+X_3^2)}},
\label{18}
 \\
J_2(0)= {{1-X_1\xi_1(0) -X_2\xi_2(0)- X_3\xi_3(0)}\over { 2(X_2^2+X_3^2)}},
\label{19}
\\
J_3(0)={{X_1\xi_1(0)-X_1^2}\over{2(X_2^2+X_3^2)}},
\label{20}
 \\
J_4(0)={{X_1(X_2\xi_3(0)-X_3\xi_2(0))}\over{2(X_2^2+X_3^2)}}.
\label{21}
\end{eqnarray}

The relations between $X_i$ and $Y_i$ values were used, because of this 
the $Y_i$-values are absent in  Eqs.(\ref{18})-(\ref{21}). Besides, we
assume that $J_{\gamma}(0)=1$. 
 Now one can obtain the differential form of these equations:
\begin{eqnarray}
{c\over \omega} {dJ_\gamma \over dx} = -J_\gamma (-{\cal{G}}
-{\cal{C}}\xi_2-{\cal{B}}\xi_3), \label{22} \\
{c\over \omega} {d\xi_1\over dx} = 
-{\cal{C}}\xi_1 \xi_2  - {\cal{B}}\xi_1 \xi_3 -{\cal{A}} \xi_2
 -{\cal{D}}\xi_3, \label{23} \\
{c\over \omega} {d\xi_2\over dx} =
{\cal{C}}(1-\xi_2^2) -{\cal{B}} \xi_2 \xi_3+{\cal{A}}\xi_1 ,  \label{24} \\ 
{c\over \omega} {d\xi_3\over dx} = 
{\cal{B}} (1- \xi_3^2)- {\cal{C}}\xi_2\xi_3 +{\cal{D}}\xi_1 . \label{25} 
\end{eqnarray}
where
\begin{eqnarray}
{\cal{A}}= (\eta'_{11} - \eta'_{22})/2,  \label{26}  \\
{\cal{B}}= (\eta''_{11}-\eta''_{22})/2,  \label{27} \\
{\cal{C}}= (\eta'_{12}-\eta'_{21})/2,    \label{28} \\
{\cal{D}}= (\eta''_{12}-\eta''_{21})/2,  \label{29} \\
{\cal{G}}=-(\tilde{n''_1}+\tilde{n''_2})= (\eta''_{11}+\eta''_{22})/2. 
\label{30}
\end{eqnarray}
It should be noted, that these equations are true only at 
$|\delta_{\alpha \beta}-\eta_{\alpha \beta}| \ll 1$ , 
where $\delta_{\alpha \beta}$ is the Kroneker 
$\delta$-function. This is due to the fact that the intensity of photon
beam is calculated approximately ($J \sim {\bf{DD}}^*$).    
 
 The differential form of equations for the general case of 
 $\eta_{\alpha \beta}$-tensor may be obtained by the similar direct
 calculations. However, another method is more conveniently to use for it.
 Knowing the relations for Stokes parameters in an 
 anisotropic medium \cite{MV,MV3},
 one can find the differential equations for this case. They 
 have a similar form as Eqs.(22)-(25). 
 Combining these equations and Eqs.(22)-(25) we get
 
\begin{eqnarray}
{c\over \omega} {dJ_\gamma \over dx} = -J_\gamma (-{\cal{G}} -{\cal{F}}\xi_1
-{\cal{C}}\xi_2-{\cal{B}}\xi_3), \label{31} \\
{c\over \omega} {d\xi_1\over dx} = 
{\cal{F}}(1-\xi_1^2) - {\cal{C}}\xi_1 \xi_2  
- {\cal{B}}\xi_1 \xi_3 - {\cal{A}} \xi_2 - {\cal{D}}\xi_3, \label{32} \\
{c\over \omega} {d\xi_2\over dx} =
{\cal{C}} (1-\xi_2^2)-{\cal{F}}\xi_1 \xi_2  
- {\cal{B}} \xi_2 \xi_3+ {\cal{A}}\xi_1 -{\cal{E}}\xi_3, \label{33} \\ 
{c\over \omega} {d\xi_3\over dx} = {\cal{B}} (1- \xi_3^2)
-{\cal{F}}\xi_1\xi_3 -{\cal{C}}\xi_2\xi_3 
+{\cal{D}}\xi_1+{\cal{E}}\xi_2, \label{34}
\end{eqnarray}
where  
\begin{eqnarray}
{\cal{E}}=(\eta'_{12} + \eta'_{21})/2 , \label{35} \\
{\cal{F}}= (\eta''_{12}+\eta''_{21})/2, \label{36}
\end{eqnarray}
and $ {\cal{A,\,B,\,C,\,D, \, G}}$ are the same as 
in Eqs.(\ref{26})-(\ref{30}). 

 It is clear that these equations are required and they describe 
 the propagation process in the general case. It obvious from following:
 
1) the form of relations for intensity and Stokes parameters  
 for the general case
 is similar to Eqs.(9)-(21). The main difference is contained that  
 $J_1(0), J_2(0), J_3(0), J_4(0)$-values have another 
 mathematical representation.
 However, it is significant that these values are linear functions
 of Stokes parameters;

 2) the derivatives of Stokes parameters in Eqs.(32)-(34) are equal to zero    
  when $\xi_1 =X_1,\,Y_1, \,\xi_2= X_2,\,Y_2$ and $\xi_3= X_3,\,Y_3$.
 This fact supports the choice of ${\cal{A}},...,\,{\cal{F}}$- parameters
 (see Eqs.(26)-(29) and (35)-(36));
 
3) Eq.(31) for intensity was obtained in \cite{MV2} for the general case. 

Note that 
$ (-{\cal{G}} -{\cal{F}}\xi_1 -{\cal{C}}\xi_2-{\cal{B}}\xi_3) \ge 0$ for
any real medium.

The obtained here equations describe the variations of intensity and  
Stokes parameters for
photon beam  propagating in the medium, whose optical properties may be
represented by the use of $\eta_{\alpha \beta}$-tensor.
Besides, one can get  the following equation for variations of
polarization $P=\sqrt{ \xi_1^2+\xi_2^2+\xi_3^2 }$ of propagating photon:
\begin{equation}
{c\over 2\omega}{dP^2\over dx}=(1-P^2)(-{\cal{F}}\xi_1-{\cal{C}}\xi_2 -
{\cal{B}}\xi_3).
\label{37}
\end{equation}
\newline 
\section{Classification}
Obtained in previous section equations describe all the cases of propagation
of photons in a medium. The difference between various cases is 
determined by the kind of $\eta_{\alpha \beta}$-tensor. 
Here we consider briefly 
all the possible cases of the photon beam propagation in media.   
 
First and foremost it should be noted that the maximal number of values,
which determinate the variations of photon state at its propagation 
in a medium, is equal to seven. However, the ${\cal{G}}$-value is
responsible only for the intensity of photon beam 
(see Eqs.(\ref{31})-(\ref{34})).
By this means only six (or less) values (${\cal{A,...,F}}$) is required 
for description of the variations of photon polarization state \cite{BF}.
On the other hand, one can make the ${\cal{E}}$-value (or ${\cal{F}}$-value)
equal to zero by rotation of the coordinate system around the wave vector of
photon. It is simplify Eqs.(\ref{31})-(\ref{34}), 
but does not reduce the number
of parameters (in this case the angle of rotation is such a parameter).
For further consideration we select coordinate system in which
${\cal{E}}=0$.

It is obviously that medium is transparent if ${dJ_\gamma \over dx} = 0$.
This condition take a place at ${\cal{G}}=0, \,{\cal{F}}=0, 
\, {\cal{C}}=0,\, {\cal{B}}=0$. The excellent example of
transparent medium is the monochromatic laser wave of not high intensity.
The beam of $\gamma$-quanta propagates in its practically without
interactions if the energy of the beam is less then the threshold of 
electron-positron pair production. Taking into account the importance
of this case for practice, we write the solution of Eqs.(\ref{31})-(\ref{34})
for transparent medium in the following form:   
\begin{eqnarray}
\xi_1(x)= \xi_1(0)\cos{\phi}+ (\xi_2(0) X_3-\xi_3(0)X_2)\sin{\phi}, 
\label{38}\\
\xi_2(x)=-\xi_1(0)X_3\sin{\phi} +X_2^2\xi_2(0) +X_2X_3\xi_3(0) +
(\xi_2(0)X^2_3 -\xi_3(0) X_2 X_3) \cos{\phi},  \label{39} \\
\xi_3(0)=\xi_1(0)X_2\sin{\phi} + X_2X_3\xi_2(0) + X_3^2\xi_3(0)+
(\xi_3(0)X_2^2-\xi_2(0)X_2X_3)\cos{\phi}, \label{40}
\end{eqnarray}
where $\phi=\Re(\tilde{n_1}-\tilde{n_2})\omega x /c$ and $X_2,\,X_3$ are
calculated with the help of Eqs.(\ref{4})-(\ref{6}).
Specifically, these equations describe the well-known cases of the light
propagation as birefringence ($X_2 =0,\,X_3=\pm 1$) and Faraday rotation
($X_2=\pm 1, X_3=0$). It easy to see that $P=const$ in the transparent 
medium. Note that the absolute transparent medium is only
physical abstraction \cite{LL}.

In general any tensor of second rank may be decomposed on the two parts:
symmetric and antisymmetric. The antisymmetric part is dual to some
axial vector, named as the vector of gyration. 
When the vector of gyration is equal to zero, the medium is named as 
anisotropic one.
It means that the optical properties of anisotropic medium are determined
by  a symmetric $\eta_{\alpha \beta}$-tensor, or, 
in other words, in the case under
consideration the following relations are true: 
${\cal{C}}=0$, ${\cal{D}}=0$. 
We remind that ${\cal{E}}=0 $ due to the choice
of the coordinate system. The anisotropic media may be classified by
the help of the  two types of $\eta_{\alpha \beta}$-tensor. In the first case
${\cal{F}}=0$, and ${\cal{F}} \ne 0$ in the second case.
The linearly polarized monochromatic
laser wave  is a simple example of the anisotropic medium  of the 
first type \cite{MV2}. 
In this case the normal electromagnetic waves are 
linearly polarized along the principal axis of the $\eta_{\alpha \beta}$-tensor
and result of propagation of the initially linearly polarized photon beam 
in this medium is the effect of birefringence. This effect applies
to transparent ${\cal{B}}=0$  and absorbed ${\cal{B}} \ne 0$ media.
However, in absorbed media of considered type \cite{MV1} 
the initially unpolarized
propagating photon beam becomes  linearly polarized 
(at $x \rightarrow \infty$). 

The case of anisotropic medium of the second type take a place 
only for absorbed media. This case was considered in different media
like single crystals \cite{MV}, plasma \cite{AG}, laser wave \cite{MV3}.
In this case the normal electromagnetic waves are elliptically polarized.  
It is interesting that $X_2=Y_2$ for this waves.
The initially unpolarized propagating in this medium
photon beam becomes linearly and circularly polarized.
The simplest  sample of the anisotropic medium of the second type 
for high energy photons
is a dichromatic laser wave. (The dichromatic wave is a superposition 
of the two linearly polarized laser waves with different frequencies
moving in the same direction and, broadly speaking, nonzero angle
direction of polarization of these wave.)  

\begin{table}[t]
\begin{center}
\begin{tabular*}{160 mm}[ ]{| r | l | l | l | l | }
\hline
$N$ & $Case$ & $Zero \,\, val.$ & $Nonzero \,\, val.$ & 
$Pol.\, states\, of\, normal\, waves$ \\
\hline
1 &$ I,\,T$ & ${\cal{A,B,C,D,F}}$ & - & Elliptically  degenerated \\
2 &$ A_1,\,T$ & ${\cal{B,C,D,F}}$ & ${\cal{A}}$ & $X_3=\pm 1, Y_3=-X_3$ \\
3 &$ A_1,\,D$ & ${\cal{C,D,F}}$ & ${\cal{A,B}}$ & $X_3=\pm 1, Y_3=-X_3$ \\
4 &$ A_2,\,D$ & ${\cal{C,D}}$ & ${\cal{A,B,F}}$ 
& $Y_2=X_2 \ne 0,Y_1=-X_1,Y_3=-X_3$ \\
5 &$ G,\,T  $ & ${\cal{A,B,C,F}}$ &${\cal{D}}$ & $X_2=\pm 1, Y_2=-X_2$ \\
6 &$ G,\,D  $ & ${\cal{A,B,F}}$ &${\cal{C,D}}$ &$ X_2=\pm 1, Y_2=-X_2$ \\
7 &$ G+A_1, T$ &${\cal{B,C,F}}$ &${\cal{A,D}}$ 
&$ X_1=Y_1=0, Y_2=-X_2\ne 0,Y_3=-X_3$ \\
8 &$ G+A_1, D$ &${\cal{F}}$ & ${\cal{A,B,C,D}}$ 
&$ Y_2=-X_2 \ne 0, Y_1=X_1, Y_3=-X_3$\\
9 &$ G+A_2, D$&- &${\cal{A,B,C,D,F}}$ 
&$ |X_2| \ne |Y_2|, X_1^2+X^2_3 \ne Y_1^2 + Y_3^2$\\
\hline
\end{tabular*}
\caption{ ${\bf{The \,\, different \,\, cases \,\, of\,\, propagation
\,\, of\,\, photons \,\, in \,\, media.}}$  
\newline
The first letter(s) in the second column is the name of case, the letter
after comma
denotes the transparent ($T$) or absorbent ($D$) medium.
Here $I$ is an isotropic medium,
$\,A_1,\,A_2$ are anisotropic media of the first and second type,
$G$ is the gyrotropic medium.${\,\cal{E}}=0$ due to the choice of
the coordinate system. Symbol $\pm$ means that one of the two signs is 
used.}
\end{center}
\end{table}

The case of isotropic medium is described by the following relations:
 ${\cal{A}}=0,\, {\cal{B}}=0,\, {\cal{C}}=0,\, {\cal{D}}=0,\, {\cal{F}}=0$.
It is obviously that  the initial polarization
state of photon beam is conserved at its propagation in the isotropic medium. 

Now we consider the medium with  the nonzero  gyration vector.
There are five cases of propagation of the photon beams.
The first and second cases are the pure (transparent and absorbent,
correspondingly) gyrotropic medium:
${\cal{A}}=0,\,{\cal{B}}=0,\, {\cal{E}}=0,\,{\cal{F}}=0$ and
one or both of the values ${\cal{C}},\,{\cal{D}}$ are not equal to zero.      
The normal electromagnetic waves are circularly polarized ($X_2=-Y_2$). 
The Faraday rotation take a place if ${\cal{C}}=0$. 
The circularly polarized monochromatic wave is the sample of such a medium
(${\cal{C}}=0$, ${\cal{D}} \ne 0$ or  both the values are not equal to zero) 
for propagating $\gamma$-beam with
energy below (above) of threshold of pair production.   

The third and fourth cases are the gyrotropic medium with anisotropy 
of the first type (${\cal{B}}=0$, ${\cal{C}}=0$, ${\cal{F}}=0$ 
for transparent
and   ${\cal{F}}=0$ for absorbent media). These cases are described with
the help of Eqs.(\ref{38})-(\ref{40}) and Eqs.(\ref{9})-(\ref{12}) 
for transparent and  absorbent  media,
correspondingly. The elliptically polarized monochromatic laser
wave is the sample of such a medium.

The last case is the gyrotropic medium with the anisotropy of the 
second type. The normal electromagnetic waves for this medium
have the different absolute values of circular and linear polarizations
($|X_2| \ne |Y_2|, X_1^2+X^3 \ne Y_1^2+Y_3^2$). The sample of this
medium is bichromatic combination of the two elliptically
polarized waves moving in the same direction with the arbitrary
angle between the semiaxis of polarization ellipses of these 
laser waves ( for propagating $\gamma$-beam above the threshold of 
pair production).   

The considered in this section cases of propagation are represented in
table 1. In this table we assume conventionally that medium is transparent
(T) independently of the quantity of  ${\cal{G}}$-value. It allows to avoid
the consideration of additional unimportant cases. 
 
Note, that the problems of  C, P, T  symmetries at propagation of the photon
beam were investigated in \cite{Co}.

\section{Discussion}
Eqs.(\ref{31})-(\ref{34}) describe  all the cases of propagation 
of photons in uniform 
(the components of $\eta_{\alpha \beta}$-tensor are constant 
along the  direction of photon beam motion at fixed $\omega$) media. 
These differential equations  of the first order have
enough a simple form and dependent on differences of the corresponding
components of $\eta_{\alpha \beta}$-tensor. 
Besides, the equations were obtained
at condition that $|\eta_{\alpha \beta} - \delta_{\alpha \beta}| \ll 1$.
One can find the some solutions of Eqs.(\ref{31})-(\ref{34})
 depending on a kind of 
$\eta_{\alpha \beta}$-tensor \cite{MV,MV2}. We do not know universal 
solution of these equations (the first integrals) 
describing  all the cases of propagation.

Eqs.(\ref{31})-(\ref{34}) one can use for calculations 
of the propagation of photon beams
in not uniform media, when the components of the $\eta_{\alpha \beta}$-tensor
are functions of coordinate along direction of the  photon wave vector.
The condition, when this approximation is warrant, is 
smallness of dimensionless parameters $c/( R \omega)$ relative to unit,
where $R$ is the distance of constancy of tensor components.
In other words, these components should be fixed on the distances of
photon wave length.   

Note, that in paper \cite{KS} the process of propagation of the
high energy $\gamma$-quanta in the monochromatic laser  wave of
arbitrary polarization was considered. For this purpose the 
well known scattering amplitudes for elastic scattering light by
light \cite{LL1} were used. As a result the differential equations
similar in form to Eqs.(\ref{22})-(\ref{25}) were obtained. 
Comparison of Eqs.(\ref{22})-(\ref{25})
and similar ones in \cite{KS}  allows to make conclusion that in this case
the components of $\eta_{\alpha \beta}$-tensor  are linear combinations
of the invariant helicity amplitudes for the forward light
by light scattering.   

The propagating in an medium  photon beam can be represented as 
a superposition of two normal electromagnetic waves with the
different refractive indices. 
 Because of this, in  absorbent media one normal wave is absorbed
to greater extent than other and after propagation of some distance
only this wave would then be left behind. 

On the other hand, every normal wave propagates in such a way that
its polarization state is conserved. Setting $d\xi_1/dx =0,\,
d\xi_2/dx=0,\, d\xi_3/dx =0$ in Eqs.(\ref{31})-(\ref{34})
 we can find the Stokes
parameters of these waves (see Eqs.(\ref{4})-(\ref{6})). 
However, in the case of transparent
medium the equations have infinite set of solutions and the third
equation is $\xi_1^2+\xi_2^2+\xi_3^2=1$.

 Eq.(\ref{37}) is the consequence of Eqs.(\ref{31})-(\ref{34}).
The total polarization $|P|$ is decreased or increased in the 
relation of sign
of the right side of Eq.(\ref{37}).  In the case, 
when ${\cal{B,\,C,\,F}}$ are equal to zero
 $P= const$. 
On the other hand, it is easy to get for the absorbent medium:  
 \begin{equation} 
 P^2(x) =1 -{{(1-P_0^2)\exp(2{\cal{G}}\omega x /c)}\over {J^2_{\gamma}(x)}}, 
 \label{41}
 \end{equation}
where $P_0 =  P(0)$ is the initial total polarization. 
Note, that $J_{\gamma}(0) = 1$. 
Besides, the value $-{\cal{G}}\omega/c$ is the inverse mean path of the  
unpolaraized photons in the medium and may be measured in principle.
The obtained simple relation is universal and may be useful for  
experimental determination of photon beam polarization.

We have considered the propagation  of monochromatic photons in media.
The absorption of photon means one of the two possibilities:
a) scattering; b) vanishing. Compton effect and pair production are
the samples  of these possibilities for $\gamma$-quanta. 
As a result, the cascade process take a place in media.
As a rule for description of cascade processes the kinetic equations are   
used \cite{HBL}. The kernels of these equations are the differential 
probabilities (in particular, per unit length) 
 of the possible elementary processes. From this point of view  
Eqs.(\ref{31})-(\ref{34}).  
may be useful in similar calculations.

We understand that our classification (see Table 1) is a matter of 
convention to a degree. For instance, cases 5 and 6 from table 1
at very small ${\cal{C}}$-value give practically the same description of
linearly polarized photon-beam propagation on short distances.
However, even so, photon beam obtains a small value of circular
polarization (in case 6). Note, that on infinity this beam becomes completely
circularly polarized, whereas in  case 5 beam remains always linearly
polarized.     
  
  Besides,  case 4 at ${\cal{A}}={\pm \cal{F}}$ and ${\cal{B}}=0$ 
may be considered as
special one. Under this condition the refractive indices of normal waves
are equal in between and $X_2=Y_2$ (see for detail \cite{AG}). However, 
we interpret its as limit of case 4.

It should be noted that the analytical solutions of 
Eqs.(\ref{31})-(\ref{34}) are exist and their full number 
is less than the number of cases  
in Table 1.  The universal solution for cases 1-3, 5-7 one can found in
the appendix of paper \cite{MV2}. The propagation in anisotropic medium
of the second type (case 4) is considered in \cite{MV,MV3}. 
Eqs.(\ref{9})-(\ref{12})  describe
the propagation process  in the gyrotropic medium with the anisotropy
of the first type. 

\section{Conclusions} 
The derived equations for intensity and Stokes parameters of
the monochromatic photons propagating in an uniform medium, whose optical
properties may be described by the permittivity tensor, is a central
result. The equations have the simple form and the variation of the
Stokes parameters is determined only by the differences of corresponding 
components of the inverse $\eta_{\alpha \beta}$ of the permittivity tensor.
It is shown that the equations may be used under certain conditions for
calculations of the photon beam propagation in not uniform media.

On the basic of these equations the classification of different cases
of photon propagation are suggested.

The problem  of generalization of Eqs.(\ref{31})-(\ref{34})
 for arbitrary quantities
of permittivity tensor components is of interest and requires  
further consideration.

Our results may be useful for investigations in astrophysics,  
and in some areas of
high energy and nuclear physics, and physics of condensed media.

\end{document}